# To what extent is researchers' data-sharing motivated by formal mechanisms of recognition and credit?

Pablo Dorta-González [1,*], Sara M. González-Betancor [2], and María Isabel Dorta-González [3]

[1] Universidad de Las Palmas de Gran Canaria, TiDES Research Institute, Campus de Tafira, 35017 Las Palmas de Gran Canaria, Spain. E-mail: pablo.dorta@ulpgc.es ORCID: http://orcid.org/0000-0003-0494-2903

[2] Universidad de Las Palmas de Gran Canaria, Departamento de Métodos Cuantitativos en Economía y Gestión, Campus de Tafira, 35017 Las Palmas de Gran Canaria, Spain. E-mail: sara.gonzalez@ulpgc.es ORCID: http://orcid.org/0000-0002-2209-1922

[3] Universidad de La Laguna, Departamento de Ingeniería Informática y de Sistemas, Avenida Astrofísico Francisco Sánchez s/n, 38271 La Laguna, Spain. E-mail: isadorta@ull.es

* Corresponding author

**Abstract**

Data sharing by researchers is a centerpiece of Open Science principles and scientific progress. For a sample of 6019 researchers, we analyze the extent/frequency of their data sharing. Specifically, the relationship with the following four variables: how much they value data citations, the extent to which their data-sharing activities are formally recognized, their perceptions of whether sufficient credit is awarded for data sharing, and the reported extent to which data citations motivate their data sharing. In addition, we analyze the extent to which researchers have reused openly accessible data, as well as how data sharing varies by professional age-cohort, and its relationship to the value they place on data citations. Furthermore, we consider most of the explanatory variables simultaneously by estimating a multiple linear regression that predicts the extent/frequency of their data sharing. We use the dataset of the State of Open Data Survey 2019 by Springer Nature and Digital Science. Results do allow us to conclude that a desire for recognition/credit is a major incentive for data sharing. Thus, the possibility of receiving data citations is highly valued when sharing data, especially among younger researchers, irrespective of the frequency with which it is practiced. Finally, the practice of data sharing was found to be more prevalent at late research career stages, despite this being when citations are less valued and have a lower motivational impact. This could be due to the fact that later-career researchers may benefit less from keeping their data private.

**1. Introduction**

Research discoveries in science are data-intensive, so data sharing by scientists is a priority in science policy (Critchlow and Van Dam, 2016). Data sharing is also a centerpiece of Open Science principles (Vicente-Sáez and Martínez-Fuentes, 2018) which, together with Open Access to publications, have the goal of making scientific research accessible to all (Dorta-González and Santana-Jiménez, 2018; González-Betancor and Dorta-González, 2019).

Citations play a vital role in open data. When researchers cite data they are making the data easier to find and putting that data at the same level as research articles in terms of importance (Martone, 2014). There is evidence of a citation advantage in sharing data. Thus, linking articles to their supporting data in a repository is associated with an increase in 25% in citations (Colavizza et al., 2020).

Citations are a measure of impact and evidence of visibility. Increasing the impact and visibility of their research is a motivating factor for researchers to share data. In this context, we use the dataset of the State of Open Data Survey 2019 by Springer Nature and Digital Science, to analyze practices of data sharing and attitudes to data citation at different stages of the research career, focusing particularly on early and late stages.

For a sample of 6,019 authors, we analyze the following research question: to what extent is researchers' data-sharing motivated by formal mechanisms of recognition and credit? For this purpose we study the relationship between the extent/frequency of the researcher's data sharing with the following four variables: how much they value data citations, the extent to which their data-sharing activities are formally recognized, their perceptions of whether sufficient credit is awarded for data sharing, and the reported extent to which data citations motivate their data sharing. Moreover, we analyze the extent to which researchers have reused openly accessible data, as well as how data sharing varies by professional age-cohort, established on the basis of the year of publication of the scientist's first peer-reviewed research article, and the relationship to the value they place on data citations.

## 2. Data sharing

Researchers recognize the benefits of data sharing (Lowndes et al., 2017) and most of them are willing to share their data in certain conditions as formal citation and open access to the results (Tenopir et al., 2011; Tenopir et al., 2020).

The attitude and practice of researchers in relation to data sharing have been analyzed in various studies (Tenopir et al., 2011; Sayogo and Pardo, 2013; Kim and Stanton, 2016; Tenopir et al., 2020). These studies focused on data sharing practices, the willingness and incentives to share, and the perceived barriers. We describe below these and other studies in relation to data sharing.

Tenopir et al. (2011) conducted a survey to 1,329 scientists. The respondents were mostly professors or lecturers (49%) and graduate or post-doctoral students (19%). The respondents' distribution was comprised of a North American majority -with 73% of the sample, 15% of participants from Europe, and the remaining participants from other regions. Those authors analyzed the data sharing practices and perceptions of the barriers and enablers of data sharing. They concluded that scientists do not make their data electronically available to others for various reasons, including insufficient time and lack of funding. Moreover, many organizations do not provide support to their researchers for data management.

Sayogo and Pardo (2013) used the same survey (Tenopir et al., 2011) and found two key determinants affecting researchers' willingness to publish their data: the data management skills and organization support, and the acknowledgement of the data set's originator.

An update of the above survey was conducted a decade later by Tenopir et al. (2020). The multinational and multidisciplinary sample of scientific researchers included 2,184 respondents. Those authors focused the analysis on examining the differences across age groups, sub-disciplines of science, and sectors of employment. Only half of the respondents were satisfied with available mechanisms for storing data. Over 85% of respondents admitted they would be willing to share their data with others and said they would use data collected by others if it could be easily accessed. A vast majority of respondents felt that the lack of access to data generated by other researchers or institutions was a major impediment to progress in science at large, yet only about a half thought that it restricted their own ability to answer scientific questions. In general, it was observed significant disciplinary and professional age-cohort differences in perceived incentives and barriers to sharing data.

Kim and Stanton (2016) used a survey to 1,317 scientists to examine to what extent institutional and individual factors influence scientists' data-sharing behaviors in a range of 43 scientific disciplines. This survey showed that regulative pressure by journals, normative pressure at a discipline level, and perceived career benefit and scholarly altruism at an individual level had significant positive relationships with data-sharing behaviors. However, perceived effort had a significant negative relationship.

There are two types of motivation for sharing data: research-intrinsic and personal (Schmidt et al., 2016). Accelerating scientific discoveries is the main research-intrinsic motivation. Personal motivations include the dissemination and recognition of research results, personal commitments to open data, and requests from data users.

Researchers are more willing to share data if directly requested to do so by another scientist, and consider that such a request will contribute to ensuring the proper use and citation of their data (Lowndes et al., 2017), as well as enhancing the possibility of being a co-author of a study that uses the data (Kim and Stanton, 2012). The type of data used and the facility with which it can reuse is another factor that influences data sharing and reuse practices (Curty et al., 2017).

Individual attitudes and practices related to data sharing should be aligned with those of funding agencies that require data sharing and institutional mandates. Studies have shown that many researchers do not share their data, even those who receive funding from agencies that require data sharing (Volk et al., 2014; Tenopir et al., 2020).

The major barriers to sharing data are related to insufficient funding and/or time to prepare data for access and reuse (Tenopir et al., 2011; Downey and Olson, 2013), concerns about the need to publish first, and legal constraints on the loss of recognition credit, especially in the early stages of the research career (Schmidt et al., 2016). In order to eliminate some of these barriers, data journals have proliferated in the last decade. The benefits of data journals are for both authors and readers. The author gets a peer-reviewed article, perhaps in a high-impact journal, and the reader obtains a dataset that has been more rigorously evaluated and more fully described than it might otherwise have been (Walters, 2020). In this respect, Candela et al. (2015) studied more than 100 data journals, and described the approaches to promote for data set description, availability, citation, quality, and open access. Buneman et al. (2020) proposed to create a different system for publishing citation summaries so that authorship could be recognized by citation analyzers. In this same aspect, Cousijn et al. (2019) described how to assess data reuse and make data usage statistics and citations available.

Legal regulations can constitute another barrier, especially in disciplines that deal with human subjects, given the possibility that shared data could be misused or misinterpreted (Downey and Olson, 2013; Schmidt et al., 2016). Researchers may also be concerned about sharing data, as the reuse of their data and the resulting scrutiny by other researchers may reveal errors/discrepancies in the datasets or in their interpretation (Gorgolewski et al., 2013).

Silvello (2018) reviewed studies on data citation from different scientific fields and analyzed why data citation is central for the development of science. He concluded that data citation is required to give credit to data creators and data curators because credit serve as an incentive to scientists for sharing more and better data, leading to the reproducibility of scientific experiments and results. He also concluded that data citation has a major impact for facilitate the discovery of data sources by providing new access points to them.

## 3. Methodology

In this study we use the State of Open Data Survey 2019 by the publisher Springer Nature and the non-for-profit technology company Digital Science. The survey respondents were reached largely via Springer Nature author lists and also distributed to the Figshare user database. The authors used the Qualtrics survey software to collect the results. The responses were collected from May to July, 2019.

This dataset contains more than eight thousand respondents, it is openly available in Nature Research et al. (2019) and described in Fane et al. (2019). After filtering by authors of at least one peer-reviewed research article among all researchers who completed more than fifty percent of the survey (including the question on data sharing frequency), we ended up with a total sample of N=6,019 researchers to conduct the present study.

Note that some respondents of the survey are users of a research resource (the open repository Figshare), which reaches out to researchers across different geographies, multiple disciplines, and at various levels of their career. However, there are implicit bias. The average Figshare user is familiar with the open research landscape, generally receptive to new technologies, new workflows, and advocates of open data (Fane et al., 2019).

For this sample of 6,019 researchers, we study:

(1) The extent/frequency of their data sharing.

(2) How much they value data citations, and the relationship with (1).

(3) The extent to which their data sharing activities are formally recognized, and the relationship with (1).

(4) Their perceptions of whether sufficient credit is awarded for data sharing, and the relationship with (1).

(5) The reported extent to which data citations motivate their data sharing, and the relationship with (1).

(6) The extent to which they have reused openly accessible data, and the relationship with (1).

(7) How data sharing varies by professional age-cohort, and the relationship with (2).

So (1) can be considered a dependent variable, in some sense, through most of the paper. And (2) becomes a dependent variable when we discuss the relationship with professional age.

Finally, we consider most of the explanatory variables simultaneously by estimating a multiple regression that predicts (1) on the basis of (2), (5), (6), and (7). Of the set of explanatory variables we had to omit (3) and (4) since the lack of response to these questions would have caused a loss of nearly 50% of the sample.

With respect to the statistical representativeness of the sample and the possible existence of survey bias, we describe below the data (with number and percentage of respondents) in relation to the field, the continent where they live, and their career status. As for the field, the respondents are from Arts & Humanities 139 (2.3%), Astronomy and planetary science 71 (1.2%), Biology 925 (15.4%), Business/Investment 109 (1.8%), Chemistry 208 (3.5%), Earth and Environmental Science 498 (8.3%), Engineering 668 (11.1%), Materials Science 133 (2.2%), Medicine 1,047 (17.4%), Physics 242 (4.0%), Social Sciences 633 (10.5%), Other 917 (15.2%), Unknown 429 (7.1%).

In relation to the continent where researchers live, respondents are from Africa 401 (6.7%), Asia -including Middle East- 1,421 (23.6%), Australasia 156 (2.6%), Europe 2,088 (34.7%), North America -including Central America and the Caribbean- 1,190 (19.8%), South America 334 (5.5%), Unknown 429 (7.1%).

With regard to the researchers' status, those surveyed are PhD/Master's student 548 (9.1%), Technician/Research Assistant 317 (5.3%), Postdoc 333 (5.5%), Healthcare professional 75 (1.2%), Research scientist 1,297 (21.5%), Professor 2,336 (38.8%), Laboratory Director/Head 248 (4.1%), Unknown 865 (14.4%).

**4. Results**

*Perceived value of data citations and relationship with data sharing frequency*

Figure 1 shows the frequency distribution of data sharing according to the perceived value of data citations. Most researchers (68.5%) value a data citation as much as a citation to an article, and the mean frequency with which they make data openly available to others is 3.20 in a Likert scale of 1 to 5. Only 19.6% of the respondents do not place much value on data citations, and the mean frequency of making data openly available to others is 2.48. Excluding the 111 respondents who do not value citations at all (1.8% of the researchers) we found that the valuation of the data citation has a considerable and positive relationship with the frequency with which scientists make data openly available to others. That is, the more scientists value data citation, the more they practice data sharing. Data citation is therefore a strong incentive for data sharing.

This relationship is verified through the non-parametric contrast of Kendall's tau for both variables, shown in Table 1, which shows a positive (0.19) and statistically significant value of Tau-b. It can also be seen in Table 2, which shows the percentage distribution of the perceived value of data citations according to the frequency of data sharing. It can be seen that the value attributed to data citation by the scientists increases with the frequency with which they share data. Most researchers (55.2%) claim to share data about or more than half the time (and 2,868 always or most of the time). The other 44.8% of the respondents share data only sometimes or never (2,698 researchers).

*Formal acknowledgement and relationship with data sharing frequency*

In relation to the formal recognition of data sharing, most respondents in the survey (58%) claim never to have received credit/acknowledgement for sharing data (Table 3), and the frequency with which those respondents make data openly available to others is 2.87 (in a Likert scale of 1 to 5). This is significantly lower than the frequency of 3.58 of the 21% of researchers who responded that they had received credit/acknowledgement for sharing data. That is, we found that credit/acknowledgement for sharing data has a notable and positive relationship with the frequency with which scientists make data openly available to others, which is verified through the non-parametric contrast of Kendall's tau, shown in Table 1, which shows a positive (0.21) and statistically significant value of Tau-b. Therefore, the formal recognition could be a strong incentive for data sharing.

To quantify the formal recognition of data sharing, we then correlated the frequency of data sharing with the scientists' perception of the credit received for this practice. Kendall's tau (Table 1) takes again a positive (0.12) and statistically significant value, showing thus a positive relation between these two variables. The percentage distribution of the scientists' perceived acknowledgement of data sharing according to the frequency with which they practice it is shown in Table 4. A small proportion of researchers (between 5.1% and 19.1%) consider that data sharing is sufficiently recognized, with this perception increasing as the frequency of data sharing grows. However, a majority of researchers (between 58.9% and 72.6%) believe that data sharing is insufficiently recognized. Distinguishing by the frequency of data sharing, that perception has an inverted u-shape. That is, the perception that data sharing is poorly recognized decreases among those who most frequently share their data. Therefore, it seems that there is a certain lack of knowledge about the actual level of recognition of this practice among some researchers.

If we consider the opinion of the more than a thousand researchers who claim that they always share their data, the proportion of those who report that it is undervalued and those who believe that it is sufficiently recognized is 3:1. That is, for each researcher who claims that data sharing is sufficiently recognized, there are three others who think it is under recognized (19.1% and 58.9%, respectively). However, within the group of researchers who practice data sharing most of the time (1,801 respondents), this ratio rises to almost 6:1 (12.8% and 69.1%, respectively).

*Data citations as motivating factor for data sharing*

The percentage distribution of the extent to which data citations are a motivation for data sharing is shown in Table 5. As can be seen, the possibility of data citations is a major motivator in data sharing, regardless of the frequency with which it is practiced. Only a small proportion of the researchers (between 7.3% and 29.6%) consider that data citations are a low or zero motivating factor for data sharing, with a higher proportion found among those who have never or only sometimes practiced data sharing, and a lower proportion among those who have always shared data or do so most of the time. Kendall's tau (Table 1) takes also a positive (0.12) and statistically significant value, showing again a positive relation between these two variables.

*Relationship between reusing open data and data sharing frequency*

The reuse of open data is an interesting factor in the decision to share data, because it is associated with previous knowledge and experience with open science. In this respect, the mean frequency of data sharing in relation to the reuse of open data is shown in Table 6. Excluding the *Don't know* respondents, about half of the researchers in the survey claim to have reused open data, and the frequency with which they make data openly available to others is 3.39 (in a Likert scale of 1 to 5), a significantly higher amount than those who have never reused open data, which is 2.74. We found that the reuse of open data has a considerable and positive relationship with the frequency with which scientists make data openly available to others, as shown through Kendall's tau (Table 1), which takes one of the highest positive significant value (0.22) of all. That is, previous knowledge and experience with open data increases data sharing, and the reuse of open data is therefore a strong incentive for data sharing.

*Professional age-cohort differences in data sharing and citation recognition*

We define the age-cohort through the year of publication of the respondent's first peer-reviewed research article. The mean frequency of data sharing by professional age-cohort is shown in Figure 2. There are considerable professional age-cohort differences in the frequency of data sharing. A u-shape can be seen in the graph and it can be observed that the most experienced researchers (seniors) share their data more frequently but that there is also a growing trend for data sharing among younger researchers (juniors).

The distribution of the year of publication of the scientists' first peer-reviewed research article in relation to the perceived value of data citations is shown in Figure 3. Again, there are considerable professional age-cohort differences in the value that researchers give to data citations. An increasing trend in the distribution by professional age is observed. That is, the higher the importance given to data citations, the lower the average professional age of the researchers. That is, younger researchers value citations more, and older researchers value citations less. This result is consistent with Kendall's Tau-b test (Table 1), which shows a positive and significant correlation (0.10) between the publication date of the first peer-reviewed article and the perceived value of data citations, implying that the younger the researcher, the greater value he or she places on data citations.

Finally, the distribution of the year of publication of the scientists' first peer-reviewed research article in relation to the extent to which receiving data citations are considered a motivating factor for data sharing is shown in Figure 4, and is tested also in Table 1 through Kendall's Tau-b test. Again, there are significant professional age-cohort differences. An increasing trend in the professional age distributions is observed. The more data citations are considered to be a motivating factor for data sharing, the lower the average professional age of the researchers. That is, younger researchers are more motivated to share data by receiving citations, and older researchers less so. Kendall's Tau-b test also shows this positive and significant correlation, although it is one of the lowest (0.03).

*Regression that estimates the extent/frequency of data sharing*

We have estimated by ordinary least-squares a regression for the extent to which authors engage in data sharing (Y) on the basis of some explanatory variables (X1 through X6) and some other control variables (C1 through C5), whose results are shown in Table 7. Of the set of explanatory variables we had to omit variables X2 (the formal acknowledgment of data sharing activities) and X3 (the researcher's perception of whether sufficient credit is awarded for data sharing), since the lack of response to these questions would have caused a loss of nearly 50% of the sample. The estimated model presents a fairly acceptable goodness of fit (about 22%), considering that it is a regression carried out with data from a cross-sectional survey, and the results obtained are consistent with the expected. Moreover, it does not present problems of heteroskedasticity or multicollinearity (see Note 5 in Table 7).

Regarding the variables of interest, it can be observed that those researchers who do not value data citations very much, compared with those who value equally article and data citations (X1), tend to be less engaged in data sharing. On the contrary, those who value data citation more than article citation tend to be more engaged in data sharing. In addition, the greater the motivation of the researcher to receive citations for their data sharing, the more engaged they are in sharing it (X4). The same applies to those who have reused open data, as opposed to those who have never done so (X5). Finally, researchers with less experience in publishing peer-reviewed articles are less likely to share data (X6).

The standardized coefficients (Beta) of these estimations allow us to determine which of these variables is greater correlated with the endogenous one. Thus, it can be seen that the variable with the greatest positive correlation with the engagement in data sharing is having ever used open data, as already indicated by Kendall's Tau-b (0.22) in Table 1. While the one that has a greater negative correlation is that of not valuing data citations very much, which is the second highest Kendall's Tau-b (0.21) in Table 1.

With regard to the control variables, it is noteworthy that the more likely the researcher is to reuse openly shared data (C1), and the greater his/her frequency of sharing data privately (C2), the greater his/her engagement in openly data sharing. Furthermore, these two are precisely the most relevant control variables, as they are the ones with the highest beta value. The other control variables show that: a) researchers working in research institutions tend to be less engaged in openly data sharing than those working for the government (C3); b) researchers in the areas of Astronomy and Planetary Science, Biology, Chemistry, and Physics tend to be more engaged in openly data sharing than researchers in the area of Medicine (C4); and c) researchers in Africa and South America tend to be more engaged in openly data sharing than researchers in North America, while the latter tend to be more engaged in openly data sharing than those in Europe (C5).

## 5. Conclusions

Science is data-intensive and data sharing is therefore an important topic in science policy and open science. In this study, we used the State of Open Data Survey 2019 (by Springer Nature and Digital Science) to research the data-sharing phenomenon, its relationship with formal mechanisms of recognition and credit, and its professional age-cohort differences. We analyzed the opinion of 6,019 researchers, all authors of at least one peer-reviewed research article.

In relation to the research question "To what extent is researchers' data-sharing motivated by formal mechanisms of recognition and credit?" results do allow us to conclude that a desire for recognition/credit is a major incentive for data sharing.

In more detail, we found that both the valuation of data citations by the researchers, as well as the credit/acknowledgement for data sharing that they receive, have a considerable and positive relationship with the frequency with which scientists make data openly available to others. That is, the data citation as a formal recognition is a strong incentive for data sharing. Moreover, we found that the possibility of receiving data citations is a great deal to data sharing, regardless of the frequency with which it is practiced.

With respect to the perceived recognition of data sharing, and considering the opinion of the more than a thousand researchers who claim that always share their data, we found that for every researcher who consider that data sharing is sufficiently recognized, there are three others who think it is under recognized. This personal perception of the low level of acknowledgment of data sharing is even more accentuated among the group of researchers who practice data sharing most of the time (for these 1,801 researchers the ratio is 6:1).

We also found that the reuse of open data has a considerable and positive relationship with the frequency with which scientists make data openly available to others. That is, previous knowledge and experience with open data increases data sharing, and reusing open data is therefore a strong incentive for data sharing.

To determine professional age-cohort differences, we used the year of publication of the scientists' first peer-reviewed research article, and found significant professional age-cohort differences in the frequency of data sharing. More experienced researchers share their data more frequently, although there is a growing trend for data sharing among younger researchers. Note there are career-stage differences in the benefits associated with keeping data private as well as career-stage differences in the benefits associated with sharing data. Later-career researchers may benefit less from keeping their data private. They have fewer years in which to benefit from projects based on privately held data, and the benefit associated with each new paper that results from the data is likely to be lower later in scholars' careers.

In addition, we found significant professional age-cohort differences in the opinion of the researchers about the importance of data citations, and the extent to which receiving data citations is a motivating factor in data sharing. Younger researchers value data citations considerable more and are more motivated to share data by the possibility of receiving data citations, whereas more experienced researchers value data citations considerable less and are less motivated to share data by the possibility of receiving data citations.

This study has possible applications in science policy and in the development of Open Science. Sharing and citing data is good for all stakeholders in Science. When researchers cite data they are making it easier to find for other scientists and are placing data at the same level as research articles in terms of importance. However, data sharing is also desirable for the researchers themselves. Citations are a measure of impact and evidence of visibility. Moreover, there is evidence of a citation advantage in sharing data.

About the limitations, note that some respondents of the survey are users of a research resource (the open data repository Figshare, owned by Digital Science), which reaches out to researchers across different geographies, multiple disciplines, and at various levels of their career. However, there are implicit bias. The average Figshare user is familiar with the open research landscape, generally receptive to new technologies, new workflows and advocates of open data.

Note that the study design does not allow us to compare formal mechanisms of recognition and credit with other possible reasons for sharing data. That is, we do not include the full range of variables that represent factors other than formal mechanisms of recognition and credit, so we cannot draw conclusions about the relative importance of those formal mechanisms. This may be a subject for further research.

The study design may also justifies the fact that our independent variables together explain just a small portion of the variation in the dependent variable (Tables 1 and 7). This is because there are some other types of motivation for sharing data.

Among the full range of variables that represent factors other than formal mechanisms of recognition and credit, we have the regulative pressure by journals, the normative pressure at a discipline level, and the scholarly altruism in accelerating scientific discoveries. All of these factors have significant positive relationships with sharing data (Kim and Stanton, 2016). Furthermore, personal motivations include also perceived career benefit as being a co-author of a study that uses the data, the dissemination of research results, personal commitments to open data, and requests from data users (Schmidt et al., 2016). Finally, the type of data used and the facility with which it can reuse is another factor that influences data sharing and reuse practices (Curty et al., 2017).

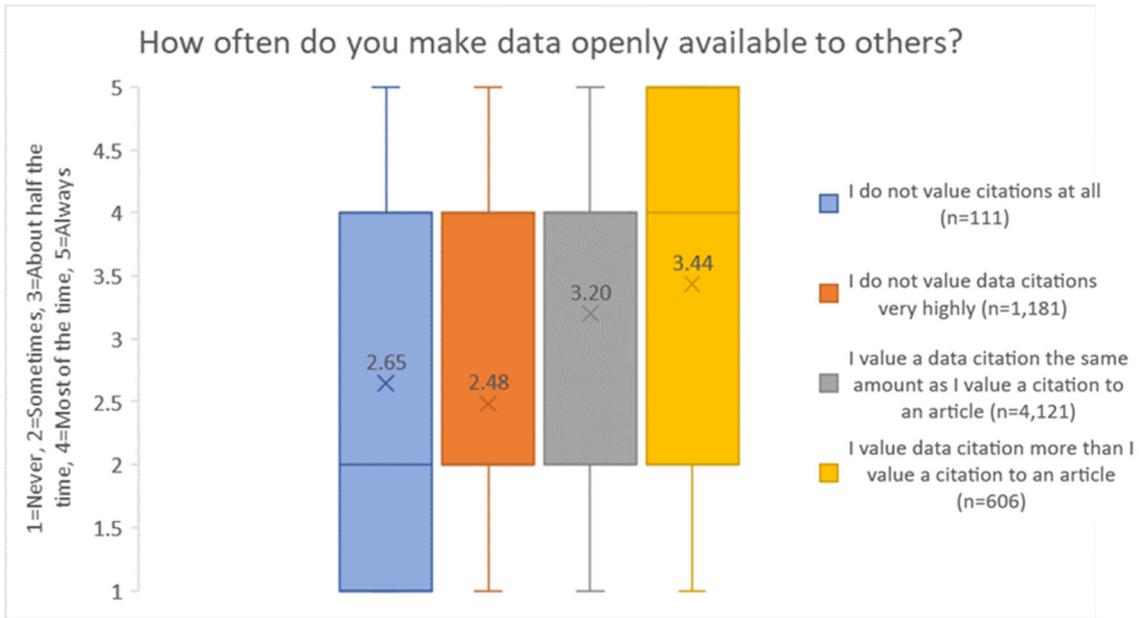

Figure 1: Box and whisker diagrams for the frequency distribution of data sharing according to the perceived value of data citations

|    | Y                    | X1                   | X2                   | X3                   | X4                   | X5                   | X6                   |
|----|----------------------|----------------------|----------------------|----------------------|----------------------|----------------------|----------------------|
| Y  |                      | 0.19** (n=6,019)     | 0.21** (n=4,734)     | 0.12** (n=4,969)     | 0.12** (n=6,019)     | 0.22** (n=5,662)     | 0.01 (n=5,604)       |
| X1 | 0.19** (n=6,019)     |                      | 0.09** (n=4,734)     | 0.06** (n=4,969)     | 0.24** (n=6,019)     | 0.10** (n=5,662)     | 0.10** (n=5,604)     |
| X2 | 0.21** (n=4,734)     | 0.09** (n=4,734)     |                      | 0.09** (n=4,000)     | 0.04** (n=4,734)     | 0.22** (n=4,517)     | -0.13** (n=4537)     |
| X3 | 0.12** (n=4,969)     | 0.06** (n=4,969)     | 0.09** (n=4,000)     |                      | -0.01 (n=4,969)      | 0.01 (n=4,708)       | 0.03* (n=4,635)      |
| X4 | 0.12** (n=6,019)     | 0.24** (n=6,019)     | 0.04** (n=4,734)     | -0.01 (n=4,969)      |                      | 0.09** (n=5,662)     | 0.09** (n=5,604)     |
| X5 | 0.22** (n=5,662)     | 0.10** (n=5,662)     | 0.22** (n=4,517)     | 0.01 (n=4,708)       | 0.09** (n=5,662)     |                      | -0.02 (n=5,278)      |
| X6 | 0.01 (n=5,604)       | 0.10** (n=5,604)     | -0.13** (n=4537)     | 0.03* (n=4,635)      | 0.09** (n=5,604)     | -0.02 (n=5,278)      |                      |

Table 1: Kendall's Tau-b correlations among all variables, with significance level and sample size (Note 1: Variable names labels are: Y= Frequency of data sharing; X1= Value of data citations; X2= Formal acknowledgment of data sharing activities; X3= Perception of whether sufficient credit is awarded for data sharing; X4= Extent to which data citations motivate data sharing; X5= Ever reused open data; X6= Year of publication of first peer-reviewed research article) (Note 2: **, and * = Statistically significant correlation at 1%, and 5% level, respectively)

|  | Frequency of data sharing | | | | |
| --- | --- | --- | --- | --- | --- |
|  | Never (n=801) | Sometimes (n=1,897) | About half the time (n=453) | Most of the time (n=1,801) | Always (n=1,067) |
| I do not value citations at all | 3.7% | 2.2% | 0.4% | 0.8% | 2.2% |
| I do not value data citations very highly | 36.0% | 25.6% | 16.8% | 11.8% | 11.2% |
| I value a data citation the same amount as I value a citation to an article | 53.1% | 64.1% | 76.2% | 76.8% | 70.5% |
| I value data citation more than I value a citation to an article | 7.2% | 8.2% | 6.6% | 10.6% | 16.1% |
|  | 100% | 100% | 100% | 100% | 100% |

Table 2: Percentage distribution of the perceived value of data citations according to the frequency of data sharing

|  | Mean frequency of data sharing (Likert 1-5) |
|---|---|
| I have never received credit/acknowledgement for sharing data (n=3,479) | 2.87 |
| I have received credit/acknowledgement for sharing data at least once (n=1,255) | 3.58 |
| Don't know / No opinion (n=1,285) | 3.11 |

Table 3: Mean frequency of data sharing in relation to the credit/acknowledgement for data sharing (Note: 1=Never, 2=Sometimes, 3=About half the time, 4=Most of the time, 5=Always)

| Frequency of data sharing | Do you think researchers currently get sufficient credit for sharing data? | | | | |
|---|---|---|---|---|---|
|  | No, they receive too little credit | Yes | No, they receive too much credit | Don't know | |
| Never (n=801) | 60.0% | 5.1% | 3.2% | 31.6% | 100% |
| Sometimes (n=1,897) | 70.0% | 8.9% | 3.3% | 17.9% | 100% |
| About half the time (n=453) | 72.6% | 9.7% | 5.5% | 12.1% | 100% |
| Most of the time (n=1,801) | 69.1% | 12.8% | 5.3% | 12.8% | 100% |
| Always (n=1,067) | 58.9% | 19.1% | 5.9% | 16.1% | 100% |

Table 4: Percentage distribution of perceived recognition of data sharing according to the frequency of its practice

|  | How much would getting data citations motivate you to make your data openly available to others? | | | | |
|---|---|---|---|---|---|
| Frequency of data sharing | None at all | A little | A moderate amount | A lot | |
| Never (n=801) | 14.1% | 15.5% | 28.1% | 42.3% | 100% |
| Sometimes (n=1,897) | 6.1% | 11.3% | 28.5% | 54.1% | 100% |
| About half the time (n=453) | 2.4% | 5.5% | 26.5% | 65.6% | 100% |
| Most of the time (n=1,801) | 3.2% | 4.1% | 19.9% | 72.8% | 100% |
| Always (n=1,067) | 7.0% | 5.2% | 16.8% | 71.0% | 100% |

Table 5: Percentage distribution of the extent to which data citations are a motivation for data sharing

|  | Mean frequency of data sharing (Likert 1-5) |
|---|---|
| I have never reused open data (n=2,844) | 2.74 |
| I have reused open data at least once (n=2,818) | 3.39 |
| Don't know (n=357) | 3.17 |

Table 6: Mean frequency of data sharing in relation to the reuse of open data (Note: 1=Never, 2=Sometimes, 3=About half the time, 4=Most of the time, 5=Always)

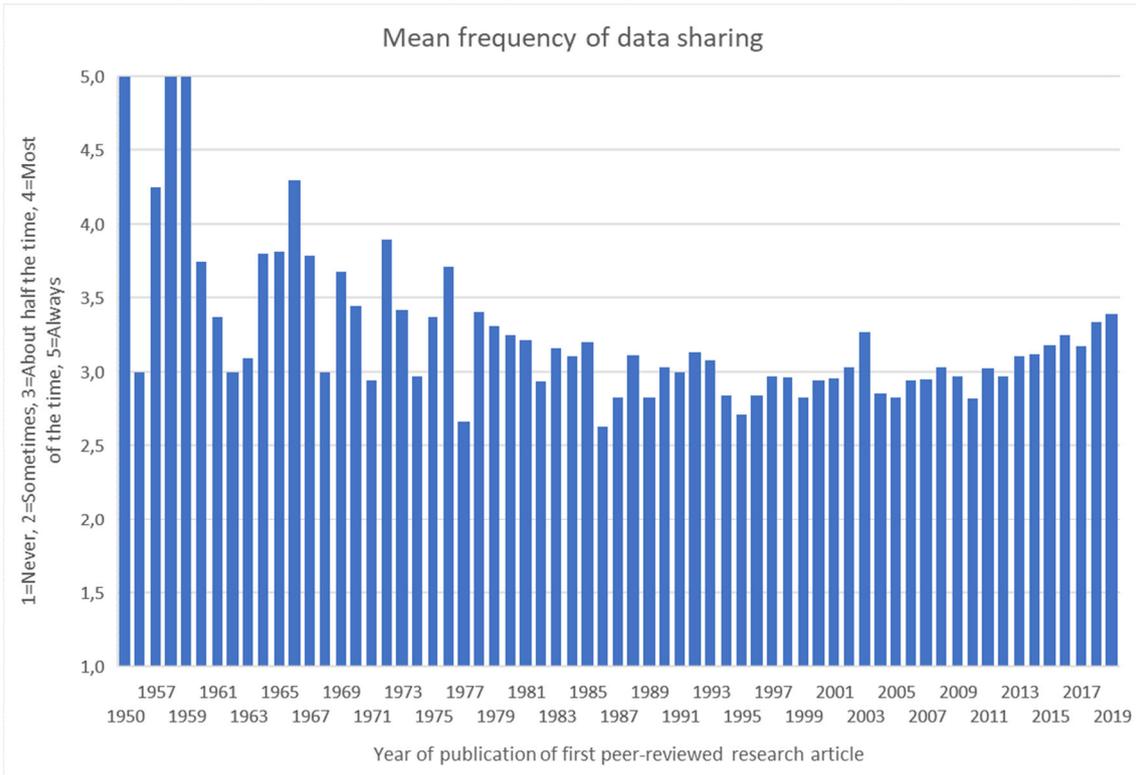

Figure 2: Mean frequency of data sharing by professional age-cohort

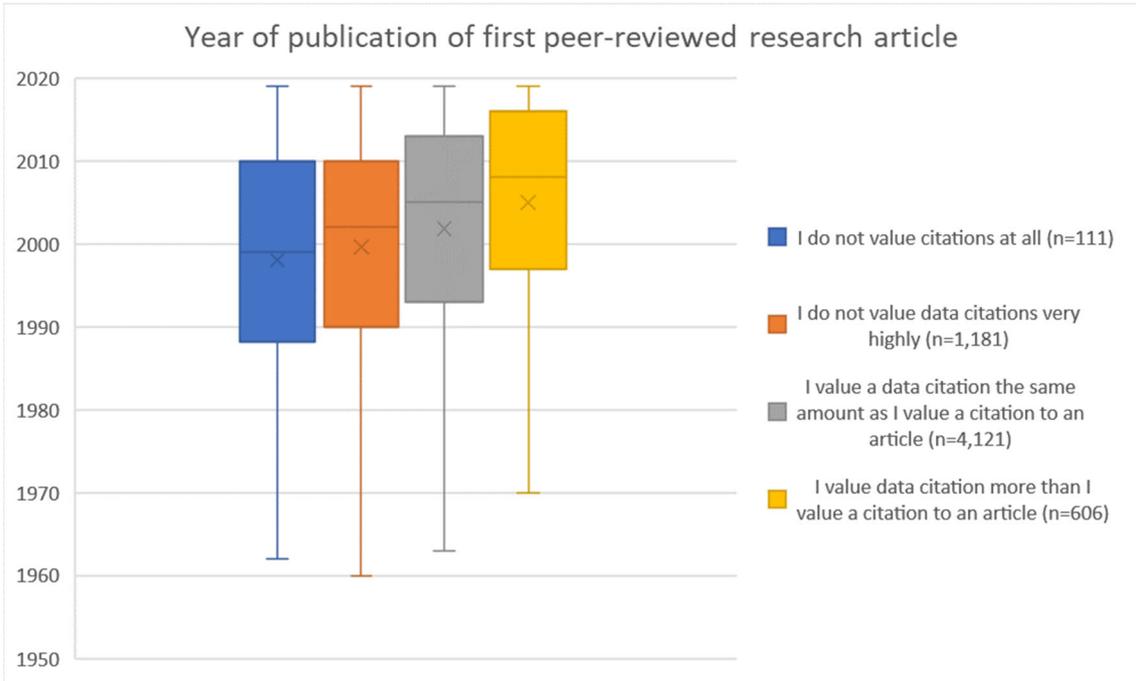

Figure 3: Box and whisker diagrams for the distribution in the year of publication of the first peer-reviewed research article according to the perceived value of data citations

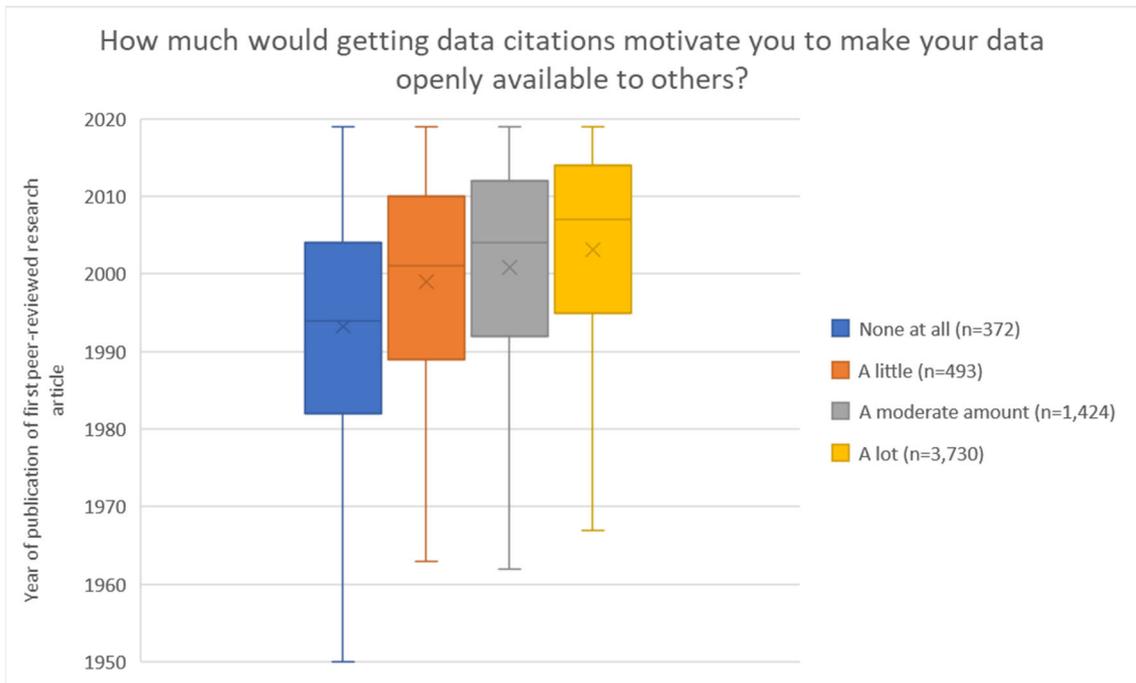

Figure 4: Box and whisker diagrams for the year of publication of the first peer-reviewed research article according to the degree of personal motivation due to receiving data citations

Table 7: Ordinary least-squares estimation for the frequency of data sharing

| Variables | Coef. | SE | Sig. | 95% C.I. | Beta |
|---|---|---|---|---|---|
| Constant | 8.938 | 2.647 | *** | (3.749, 14.127) | . |
| **X1: Value of data citations** | | | | | |
|    Values data citation equally to article citation | (Ref) | | | | |
|    Does not value citations at all | -0.135 | 0.128 | | (-0.386, 0.115) | -0.013 |
|    Does not value data citations very high | -0.426 | 0.045 | *** | (-0.514, -0.336) | -0.125 |
|    Values data citation more than article citation | 0.187 | 0.057 | *** | (0.074, 0.298) | 0.041 |
| **X4: Extent to which data citations motivate data sharing** | 0.052 | 0.016 | *** | (0.021, 0.082) | 0.045 |
| **X5: Ever reused open data** | | | | | |
|    No | (Ref) | | | | |
|    Yes | 0.386 | 0.037 | *** | (0.313, 0.458) | 0.142 |
| **X6: Year of first peer-reviewed research article** | -0.004 | 0.001 | *** | (-0.006, -0.001) | -0.039 |
| C1: Likelihood of open data reuse | 0.204 | 0.018 | *** | (0.167, 0.240) | 0.156 |
| C2: Frequency of making data privately available | 0.232 | 0.011 | *** | (0.210, 0.254) | 0.254 |
| C3: Type of working organization | | | | | |
|    Government/local government | (Ref) | | | | |
|    Hospital | -0.075 | 0.121 | | (-0.312, 0.162) | -0.011 |
|    Medical School | -0.123 | 0.113 | | (-0.343, 0.097) | -0.020 |
|    Private company | -0.100 | 0.118 | | (-0.330, 0.130) | -0.014 |
|    Research institution | -0.275 | 0.107 | ** | (-0.484, -0.065) | -0.046 |
|    University | -0.096 | 0.092 | | (-0.276, 0.084) | -0.025 |
|    Other (please specify) | -0.106 | 0.083 | | (-0.269, 0.057) | -0.038 |
| C4: Area of interest | | | | | |
|    Medicine | (Ref) | | | | |
|    Arts & Humanities | 0.130 | 0.119 | | (-0.103, 0.363) | 0.014 |
|    Astronomy and Planetary Science | 0.350 | 0.155 | ** | (0.046, 0.653) | 0.029 |
|    Biology | 0.273 | 0.061 | *** | (0.153, 0.392) | 0.075 |
|    Business/Investment | -0.128 | 0.127 | | (-0.376, 0.121) | -0.013 |
|    Chemistry | 0.168 | 0.098 | * | (-0.024, 0.360) | 0.023 |
|    Earth and Environmental Science | 0.045 | 0.073 | | (-0.097, 0.187) | 0.010 |
|    Engineering | -0.045 | 0.068 | | (-0.177, 0.088) | -0.011 |
|    Materials Science | -0.083 | 0.119 | | (-0.315, 0.149) | -0.009 |
|    Physics | 0.192 | 0.061 | *** | (0.071, 0.312) | 0.052 |
|    Social Sciences | 0.082 | 0.094 | | (-0.102, 0.265) | 0.012 |
|    Other (please specify) | -0.176 | 0.068 | *** | (-0.309, -0.043) | -0.041 |
| C5: Living continent | | | | | |
|    North America (including Central America) | (Ref) | | | | |
|    Africa | 0.101 | 0.074 | | (-0.044, 0.246) | 0.019 |
|    Asia (including Middle East) | 0.087 | 0.052 | * | (-0.016, 0.189) | 0.028 |
|    Australasia | -0.116 | 0.106 | | (-0.324, 0.091) | -0.014 |
|    Europe | -0.110 | 0.047 | ** | (-0.200, -0.018) | -0.039 |
|    South America | 0.202 | 0.078 | *** | (0.050, 0.354) | 0.036 |

Note 1: The variables of interest are in the unshaded rows. In the shaded rows are the control variables.

Note 2: Sample size and Goodness of fit measures: N = 5266; $R^2$ = 0.222; $R^2$ Adjusted = 0.218; $F(30,5235)$ = 49.84; Prob > F = 0.000; RMSE = 1.201

Note 3: ***, **, and * = Statistically significant coefficient at 1%, 5%, and 10% level, respectively.

Note 4: Reference: Researcher who values data citation equally to article citation, who never has reused open data, who works for the Government, whose area of interest is Medicine and lives in North America.

Note 5: Post estimation test for heteroskedasticity of Breusch-Pagan / Cook-Weisberg: $Chi^2_1$=0.85 (p=0.3564); Mean VIF = 1.69